\documentclass[12pt,a4paper,final]{iopart}

\newcommand{\od}{{\rm d}}
\newcommand{\omits}[1]{}
\usepackage{iopams}
\expandafter\let\csname equation*\endcsname\relax

\expandafter\let\csname endequation*\endcsname\relax

\usepackage{amsmath}
\usepackage{graphicx}
\usepackage[breaklinks=true,colorlinks=true,linkcolor=blue,urlcolor=blue,citecolor=blue]{hyperref}
\usepackage{bm}
\begin{document}

\title[]{Motion of photons in a background of gravitational wave}

\author{Zhe Chang, Chao-Guang Huang, and Zhi-Chao Zhao}
\address{Institute of High Energy Physics, \\ Chinese Academy of
Sciences, Beijing 100049, China \\
School of Physics, University of Chinese Academy of Sciences, Beijing 100049. China}

\begin{abstract}
The photon motion in a Michelson interferometer is re-analyzed in both geometrical optics
and wave optics.  The classical paths of the photons in the background of gravitational
wave are derived from Fermat principle, which is the same as the null geodesics in general relativity.
The deformed Maxwell equations and the wave equations of electric fields in the background of gravitational wave are presented in flat-space approximation.
Both methods show that the response of an interferometer depends on the frequency of a
gravitational wave, however it is almost independent of the frequency of the mirror's vibrations.
It implies that the vibrating mirror cannot mimic a gravitational wave very well.
\end{abstract}

\vspace{2pc}

\vspace{2pc}
\submitto{\CQG}

\section{\label{sec:level1}Introduction}

The transient gravitational wave (GW) events, GW150914 and GW151226,
were detected in the first run (O1, from September 12, 2015 to January 19, 2016)
of the advanced LIGO
{\cite{LVC00}, \cite{LVC10}}.  The events were reported to be the results of the
coalescence of two black holes of masses $36 M_\odot$and $29 M_\odot$ to a black hole
of $62 M_\odot$ and of two black holes of masses about $14 M_\odot$ and
$8 M_\odot$ to a black hole of mass about $21 M_\odot$, respectively.  The GW
signals are chirps in the frequency ranges between 35 to 150 Hz and
between 35 to about 700 Hz, respectively.  The signals of GW150914 and
GW151226 are extracted from the sea of noises by the same methods (for example,
see refs. \cite{Black}, \cite{MatchFilter}).  Hence, the explanation of the observations  are widely accepted.

A LIGO detector is, basically, a Michelson intermerometer (MI) with its arms replaced by
Fabry-P\'erot (FP) cavities \cite{aLIGO}.  One of the roles of the FP cavities is
to increase the interaction time of photons with GWs.  On average,
a photon travels 140 round-trips in the FP cavity.  In other words, a photon stays in
a cavity for about $280L/c \approx 3.73$ ms on average, where $L\approx 4$ km is the average
length of arms.  In 3.73 ms, a GW with frequency about 268 Hz will propagate through
the detector for a wavelength and thus a photon cannot `feel' the length variation of the
arms \cite{SC}, \cite{CPC}.  However, it is widely accepted that the response of a Michelson-Fabry-P\'erot interferometer is as about $2N$ times as the response of a same size
Michelson interferometer, where $N$ is the average number of the round trips (see, for example, \cite{MRSWD},\cite{Black}).  The previous studies on the instruments and calibration \cite{MRSWD}, \cite{DynRes}, \cite{Readout}, \cite{Character}, \cite{TestMass}, \cite{cqg}, \cite{calibration} show that the response of a
detector to a GW will not depend on the frequency when the frequency of a GW is lower than 1 kHz.
Why is there such a kind of inconsistency?

In fact, almost all previous studies on the response of detector to a GW are conducted
by use of the swing of mirrors \cite{DynRes}, \cite{Readout}, \cite{Character}, \cite{cqg}, \cite{calibration}.  They seem to be supported by the theoretical analysis for the response to
a GW \cite{TestMass}.  In these analysis, the Laplace transformation is
used, which is still questionable\footnote{To our knowledge, the first application of the Laplace transformation in the study of
the response of an interferometer is in \cite{Mizuno} and \cite{MRSWD}.  In the references, the following equation is used
\begin{equation*}
  {\cal L}\left (\int_{t-L}^th(t)dt\right ) = \frac {1-e^{-pL}}{p}{\cal L}(h(t)),
\end{equation*}
where ${\cal L}$ represents the Laplace transformation, $p$ is the Laplace domain variable.
However, for $h(t)=\sin (\omega t)$, a direct calculation shows
\begin{equation*}
  {\cal L}\left (\int_{t-L}^t \sin(\omega t)dt\right ) =
  {\cal L}\left (\frac 1 \omega \left(\cos(\omega (t-L))-\cos(\omega t)\right)  \right ) \\
  =\frac 1 \omega \frac 1 {p^2+\omega^2}(p\cos(\omega L)+\omega\sin(\omega L)-p),
\end{equation*}
and
\begin{equation*}
  \frac {1-e^{-pL}}{p}{\cal L}(\sin(\omega t)) = \frac {1-e^{-pL}}{p} \frac{\omega} {p^2+\omega^2}.
\end{equation*}
They are not equal to each other.  The equation should be revised as
\begin{equation*}
  {\cal L}\left (\int_{t-L}^th(t)dt\right ) = \frac {1-e^{-pL}}{p}{\cal L}(h(t))-e^{-pL}
  \left [\int_{-L}^0 e^{-p\tau}\left( \int_0^\tau h(t)\od t\right )\od\tau\right].
\end{equation*}}.
In the present paper, we shall re-analysis the response of a detector to a GW, without the help of Laplace transformation. We shall focus on the Michelson
interferometer in this paper.

The motion of photons in a detector is govern by general relativity.
In other words, the photons should move along the null geodesics in the
space-time.  In fact, there are two other ways to describe the motion of photons.
The first one is the method of classical geometrical optics, in which
paths of photons in three dimensional space are determined by the Fermat
principle.  To apply the Fermat principle in vacuum of a curved spacetime,
the metric is regarded as a position-, direction-, and time-dependent refractive-index tensor
of a special transparent medium in a flat space.
The second one is the method of classical wave optics, which
is based on the wave equation of electric field derived from the Maxwell equations.
The first aim of the present paper is to study the motion of photons in a single arm
by use of the two methods. Although a realistic detector needs double arms, we may just consider a single one because the motion of photons in each arm is govern by the same law but has the different polarity due to the quadruple nature of the GW.  We shall show that the paths of photons determined by the
Fermat principle are the same as that determined by the null geodesics even
in the background of GW.  We shall also show that a GW may serve as
a time-dependent medium in flat space-time and present the deformed
Maxwell equations and wave equations of electric field in the flat space-time
approximation. With the two methods, we may obtain the same claim as that by use of
the null geodesics \cite{SC}, \cite{CPC}:  the response
of a detector depends on the frequency of a GW.

The key point for an interferometric detector of GW is to measure the variation of the
phase difference between two beams of light.  The variation of the phase difference comes
from the variation of proper lengths of two arms induced by a GW.
On one hand, GW signals are submerged in the sea of noises and one has to find ways
to extract the GW signals from noises.  On the other, a detector needs the
calibration to determine the sensitivity of the detector.
Since both a GW and the mirror's swings will result in the variations of the proper
length of each arm, a naive idea is that the effect of GW on a detector can be mimicked by the vibrations of the mirrors' position and many detector investigations have been
made in this way \cite{DynRes}, \cite{Readout}, \cite{Character}, \cite{cqg}, \cite{calibration}.  The other purpose of the paper is
to show that there is essential difference for the motion of photons in the background of
GW and in a vibrating arm.  The response of a detector depends on the
frequency of a GW.  For some specific frequency, the detector has no response.  In contrast, the response of a detector to the vibrations with different frequency is almost the same.

The rest of the paper is organized as follows. In section 2, the motion of photons in
both a flat space-time and in the background of GW is studied in
the framework of geometrical optics.  The trajectories of photons are derived from the
Fermat principle and are shown to be the same as those along null geodesics
in general relativity.  The time difference for a round trip between in the background of GW
and in a flat space-time is shown to be dependent on the frequency of GW.
In section 3, the propagation of the electromagnetic wave is investigated
in wave optics.  The deformed Maxwell equations and thus the wave equations for electric
field are presented in the background of GW in a flat space-time approximation. The time difference is also calculated in the framework of wave optics.  The results
are the same as those in geometrical optics.  Section 4 is devoted to the comparison of the differences between motion of photons in the background of GW and in a vibrating arm.
The conclusions and remarks are given in section 5.  Throughout the paper, the $c=1$ units are used in the derivation and thus $x^0=t$.

\section{\label{sec:level1}Motion of photons in the background of gravitational wave}

We are interested in the Minkowski spacetime and the spacetime with a plus-polarized, linear, normally incident, plane GW.
In a time-orthogonal coordinate system, the metrics for the both cases can be
written as
\begin{equation}\label{1}
 \od s^2=-\od t^2+g_{ij}\od x^i\od x^j,~i,j=1,2,3~.
\end{equation}
The distance between two near points in a simultaneous hypersurface is measured by \cite{Landau}
\begin{equation}\label{eq:distance}
  \od l^2=g_{ij}\od x^i \od x^j.
\end{equation}
We consider  a photon travels from point $A$ to point $B$ in space. There
are infinitely possible classical paths.  The length for each path is given by
\begin{equation}\label{2}
  L=\int_{A}^{B}\left (g_{ij}\frac{\od x^i}{\od t}\frac{\od x^j}{\od t}\right )^{1/2}\od t,
\end{equation}
where ${\od x^i}/{\od t}$ is the coordinate velocity of the photon.

In geometrical optics, the path of a photon is determined by the Fermat principle,

\begin{align}
  0 & = \delta L  = \delta \int_{A}^{B}\left (g_{ij}\frac{\od x^i}{\od t}\frac{\od x^j}{\od t}\right )^{1/2}\od t  \notag \\
  & = \frac 1 2 \int_{A}^{B} \dfrac{\delta g_{ij}\dfrac{\od x^i}{\od t}\dfrac{\od x^j}{\od t}+2g_{ij}\dfrac{\od\delta x^i}{\od t}\dfrac{\od x^j}{\od t}}{\sqrt{g_{mn}\dfrac{\od x^m}{\od t}\dfrac{\od x^n}{\od t}}}\od t \notag \\
  & = \int_{A}^{B} \left[\dfrac{g_{ij,k}\delta x^{k} \dfrac{\od x^i}{dt}\dfrac{\od x^j}{\od t}}{\sqrt{g_{mn}\dfrac{\od x^m}{\od t}
   \dfrac{\od x^n}{\od t}}} - 2\delta x^i\dfrac{\od}{\od t}\left(\dfrac{g_{ij}\dfrac{\od x^j}{\od t}}
   {\sqrt{g_{mn}\dfrac{\od x^m}{dt}\dfrac{\od x^n}{\od t}}}\right)\right]\od t~.
\end{align}
Here, only the isochronal variation is considered, so that $\delta t =0$.
In the last step, the total-derivative term is absent because the end point of the integral
is fixed.
From the above expression, it is easy to see that
\begin{align} \label{5}
  \dfrac{ g_{ij,k}\dfrac{\od x^i}{\od t}\dfrac{\od x^j}{\od t}}{\sqrt{g_{mn}\dfrac{\od x^m}{\od t}
   \dfrac{\od x^n}{\od t}}}-2\dfrac{\od }{\od t}\left(\dfrac{g_{k j}\dfrac{\od x^j}{\od t}}
   {\sqrt{g_{mn}\dfrac{\od x^m}{\od t}\dfrac{\od x^n}{\od t}}}\right)=0~.
\end{align}
For convenience, we take $l$ as the path parameter.
Then, Eq.(\ref{5}) becomes
\begin{align}
   g_{ij,k}\dfrac{\od x^i}{\od l}\dfrac{\od x^j}{\od l}-2\dfrac{\od }{\od l}\left(g_{k j}\dfrac{\od x^j}{\od l}\right)=0~,
\end{align}
or
\begin{align}
  \dfrac{1}{2}g^{j \rho}\left(g_{\mu \nu, \rho}-2g_{\rho \mu, \nu}\right)\dfrac{\od x^\mu}{\od l}\dfrac{\od x^\nu}{\od l}-\dfrac{\od ^2x^j}{\od l^2}=0~.
\end{align}
In terms of the Christoffel symbols, it can be recast into the spatial
part of 4-dimensional geodesic equation:
\begin{equation}\label{geodesic}
  \dfrac{\od^2x^i}{\od l^2}+\Gamma^i_{\mu
  \nu}\dfrac{\od x^\mu}{\od l}\frac{\od x^\nu}{\od l}=0,~\mu ,\nu =0,1,2,3~.
\end{equation}
It should be noted that Eq. \eqref{geodesic} is different, in general, from the geodesic equation in 3-dimensinal space,
\begin{equation}\label{3d-geod}
  \dfrac{\od^2x^i}{\od l^2}+\Gamma^i_{jk}\dfrac{\od x^j}{\od l}\frac{\od x^k}{\od l}=0.
\end{equation}
The difference comes from the time-dependence of $g_{ij}$ in \eqref{geodesic} or, in other words, from the time-dependent `refractive index'.

Since the light travels along a null curve and since the path length should take
extremal value, the traveling time of a photon from $A$ to $B$
\begin{equation}\label{eq:null}
  T=\int_A^B \od t =\int_A^B \left(g_{ij}\dfrac{ \od x^i}{\od t}\dfrac{ \od x^j}{\od t}\right )^{-1/2}\od l
\end{equation}
should also take the extremal value for fixed path. Namely,
\begin{eqnarray}
0&=&  \delta T =-\frac 1 2 \int_A^B\left (g_{mn}\frac{\od x^m}{\od t}\frac{\od x^n}{\od t} \right )^{-3/2}\notag \\
  &&\qquad \qquad \left[ \delta g_{ij}\frac{\od x^i}{\od t}\frac{\od x^j}{\od t}-2g_{ij}\frac{\od x^i}{\od t}\frac{\od x^j}{\od t}\frac{\od\delta t}{\od t}  \right ] \od l \notag \\
 &=&-\frac 1 2 \int_A^B\left[\left (g_{mn}\frac{\od x^m}{\od t}\frac{\od x^n}{\od t} \right )^{-1/2}g_{ij,0} \frac{\od x^i}{\od l}\frac{\od x^j}{\od l} \right . \notag \\
  &&\qquad \qquad \left . +2\frac{\od}{\od t}\left (g_{ij}\frac{\od x^i}{\od t}\frac{\od x^j}{\od t} \right )^{-1/2}  \right ]\delta t \od l \notag \\
  &=&-\frac 1 2 \int_A^B\left (g_{mn}\frac{\od x^m}{\od t}\frac{\od x^n}{\od t} \right )^{-1/2}\left[g_{ij,0} \frac{\od x^i}{\od l}\frac{\od x^j}{\od l} \right . \notag \\
  &&\qquad \qquad \left . +2 \left (g_{pq}\frac{\od x^p}{\od t}\frac{\od x^q}{\od t} \right )\frac{\od}{\od l}\frac{\od t}{\od l} \right ]\delta t \od l .
\end{eqnarray}
Here, only the time variation is considered ($\delta x^i=0$).  In the second step, the total derivative is zero, too.  Hence,
\begin{equation}\label{pre-time-geod}
  g_{ij,0} \frac{\od x^i}{\od l}\frac{\od x^j}{\od l} +2\left (\frac{\od l}{\od t} \right )^2\frac{\od}{\od l}\frac{\od t}{\od l}=0.
\end{equation}
When
\begin{equation}\label{null}
  \left (\frac{\od l}{\od t} \right )^2=1 ,
\end{equation}
in terms of the Christoffel symbols, Eq. \eqref{pre-time-geod} reduces to the temporal part of four-dimensional geodesic equation:
\begin{equation}\label{geodesic-t}
  \dfrac{\od^2 t}{\od l^2}+\Gamma^0_{\mu
  \nu}\dfrac{\od x^\mu}{\od l}\frac{\od x^\nu}{\od l}=0~.
\end{equation}

Eqs. \eqref{geodesic}, \eqref{geodesic-t}, \eqref{null} show that in the framework of the geometrical optics a photon travels along a null geodesic of a 4-dimensional space-time in both cases of a flat space-time and the
background of GW, as expected in general relativity and the arc length $l$ along a path
serves as the affine parameter of the null geodesic.  The trajectories of photons are the
straight line in the 3 space just as the absence of a GW.  The distance that
a photon travels is the arc-length of the geodesic, $ L = \int \od l.$   Its value and thus
the traveling time will be influenced by the time-dependent metric $g_{\mu \nu}$ in the
background of GW.

Now, we calculate the traveling time of a photon travels along $x$ axis from
mirror $A$ at position $x_1=0$ to mirror $B$ at $x_2=L$.  (From now on, $L$ is the
coordinate length of the arm.)
When there is no GW and matter, the metric is of the form $g_{\mu\nu}={\rm diag}(-1, 1, 1, 1)$.
When a monochromatic GW with plus-polarization incidents along the $z$ axis,
the metric has the form
\begin{equation}\label{metric2}
  g_{\mu\nu}={\rm diag}(-1, 1+h_{11}, 1-h_{11}, 1), ~~~
  h_{11}=\Re [e_{11}e^{-i (2\pi f t+\phi)}],
\end{equation}
where $e_{11}$ is the amplitude of the GW, $\Re$ means that the real part
of the expression is taken, $\phi$ is the initial phase of the GW when the photon leaves
mirror $A$.

It is well known that if there is no GW, the
distance between two mirrors in one arm of the interferometer
is
\begin{equation}\label{10}
  L=\int_{0}^{L} \od x .
\end{equation}
It is also the traveling time of a photon from one mirror to the other.
In other words, $T=L$.

In the background of GW, the traveling time of a photon
from one mirror to the other becomes $T'$ and
\begin{align}\label{11}
  T'  = \int_{0}^{L} (g_{11})^{-1/2} \od x = \int_{0}^{L} \left(1+\frac{1}{2}h_{11}\right)\od x~.
\end{align}
In the last step of \eqref{11}, the Taylor expansion has been used and
the GW wave is supposed to be an extremely small perturbation around
the flat space.   Obviously,
\begin{equation}
  T'= L + \frac{1}{2}\int_{0}^{L}h_{11}\left(t\right)\od x=T + \Delta T'~.
\end{equation}
As expected, a GW background will change the length
of the interferometer's arm.  In other words, a photon will `feel' the distance difference
$\Delta L= \Delta T' =(1/2) \int_{0}^{L}h_{11}(t)\od x$ due to the presence of a GW.
One should remember that $h_{11}$ is a function of time $t$.

Now we consider a photon moves a round trip from mirror $A$ to mirror
$B$ and then back to mirror $A$. For $A \rightarrow B$, the distance that
a photon travels is
\begin{align}
  T'_{A \rightarrow B} =T+\Delta T'_{A\rightarrow B}~,
\end{align}
and for $B \rightarrow A$, the distance that a photon travels is
\begin{align}
  T'_{B\rightarrow A} & =\int_{L}^{0}
  \left(1+\frac{1}{2}h_{11}\left(t\right)\right)\od x = L + \frac{1}{2}\int_{t_0+T}^{t_0+2T}h_{11}\left(t\right)\od t
    = T+\Delta T'_{B\rightarrow A},
\end{align}
where $t_0$ is the time when the photon leaves $A$.
In the first line of the equation, d$x$ is always negative.  In the second line,
the higher-order terms have been neglected.
It should be noted that $\Delta T'_{A \rightarrow B}\ne \Delta T'_{B \rightarrow A}$
in principle.

For a round trip, we have
\begin{align}\label{13}
  \Delta T' & = \frac{1}{2}\int_{t_0}^{t_0+2T}
h_{11} \left(t\right)\od t 
    = \frac{e_{11}\sin(2\pi f T)}{2\pi f}\Re\left[e^{-i
   (\phi+2\pi f (t_0+T))}\right ]~.
\end{align}
It shows that at specific values of the frequency of
GW, $f=\frac{n}{2L}$ (with $n=1,2,3,...$), $\Delta T' =0$ no matter what
value the initial phase $\phi$ takes.  It means
that for the GW with the frequency $\frac{n}{2L}$
there will be no measure difference between the presence and absence of
a GW for traveling photons.  The result has been reached in the literature \cite{Thorne}, \cite{MRSWD}.

Define the dimensionless frequency of GW: $a=2Lf (>0)$.  Then,
the traveling time of a photon for a round trip from $A$ to $B$ and back to $A$ is
\begin{equation}
  \Delta T'=\frac{e_{11}L}{2\pi a}\Re\left[-i e^{-i\left(\phi+\frac{a\pi t_0}{L}\right)}
  \left(1-e^{-2i a \pi}\right)\right]~.
\end{equation}
In particular, when $t_0=0$,
 \begin{equation}\label{3.4}
  \Delta T'=\frac{e_{11}L}{2\pi a}\left[\sin\left(2a\pi +\phi\right)-\sin
  \left(\phi\right)\right]~.
\end{equation}

\section{\label{sec:level1}Propagation of electromagnetic wave in the background of gravitational wave}

In wave optics, light is treated as electromagnetic wave, which satisfies the Maxwell
equations.

The Maxwell equations in vacuum of a curved space-time read
\begin{equation}\label{maxequation1}
F^{\mu \nu}_{;\nu}=\frac{1}{\sqrt{-g}}
\left(\sqrt{-g}F^{\mu \nu}\right)_{,\nu}=0~,
\end{equation}
\begin{equation}\label{maxequation2}
F_{\mu \nu , \lambda}+F_{\lambda \mu , \nu}+F_{\nu \lambda ,
  \mu}~=0~,\
\end{equation}
where $F_{\mu \nu}=-F_{\nu \mu} = \partial_{\mu} A_{\nu} - \partial_{\nu} A_{\mu} \,$
is the electromagnetic field tensor, $A_\mu$ is the electromagnetic potential.
The electric field is
\begin{equation}\label{E}
  E_i=-F_{i\nu}U^{\nu}~,
\end{equation}
and the magnetic field is
\begin{equation}\label{B}
 B_{i}=-\frac{1}{2}\epsilon_{i\mu}{}^{\lambda \sigma}U^{\mu}F_{\lambda \sigma }~,
\end{equation}
where $\epsilon$ is the Levi-Civita tensor and $U^\mu$ is the 4-velocity of an observer,
satisfying $g_{\mu \nu}U^\mu U^\nu=-1$.  For static observers, $U^\mu=(1,0,0,0)$.
Then, the electric field $E$ and magnetic field $B$ observed by the static observers are
\begin{equation}\label{Ei}
    E_i=-F_{i\nu}U^\nu=-F_{i0}=\partial_0A_i-\partial_iA_0~,
\end{equation}

\begin{equation}\label{Bi}
    B_i=-\frac{1}{2}\epsilon_{i\mu}^{\ \ \lambda \sigma}U^{\mu}F_{\lambda \sigma }
    =\frac{1}{2}\epsilon_{i}^{\ \lambda \sigma}F_{\lambda \sigma } \\
    =\frac{1}{2}
    g^{\alpha \lambda}g^{\beta \sigma}\epsilon_{i \alpha \beta }
    \left(\partial_\lambda A_\sigma-\partial_\sigma A_\lambda\right)~.
\end{equation}

In the background of GW (\ref{metric2}), the explicit forms of three components
of the magnetic filed $B_i$ are,
\begin{equation}
     \begin{array}{ll}
       B_1=
       \left(1+h_+\right)
       \left(\partial_2 A_3-\partial_3 A_2\right)~,  \\
       B_2=\left(1-h_+\right)
       \left(\partial_3 A_1-\partial_1 A_3\right)~,  \\
       B_3=
       \left(\partial_1 A_2-\partial_2 A_1\right)~.
     \end{array}
\end{equation}
Therefore,
\begin{equation}
       \begin{array}{ll}
      F_{0i}=E_i~,  \\
      F_{12}=B_3~,   \\
      F_{23}=\left(1-h_+\right)B_1~,  \\
      F_{31}=
      \left(1+h_+\right)B_2~,
    \end{array}
   \end{equation}
and,
\begin{equation}
      \begin{array}{ll}
      F^{01}=-

      \left(1-h_+\right)E_1~,  \\
      F^{02}=-\left(1+h_+\right)E_2~,   \\
      F^{03}=-E_3~,
    \end{array}
   \end{equation}
\begin{equation}
       \begin{array}{ll}
      F^{12}=B_3~,   \\
      F^{23}=B_1~,   \\
      F^{31}=B_2~.
    \end{array}
   \end{equation}

In the background of GW (\ref{metric2}), the Maxwell equations \eqref{maxequation1} read
\begin{equation}\label{mx1}
  F^{\mu 0}{}_{,0}+F^{\mu i}{}_{,i}=0~.
\end{equation}
Namely,
\begin{equation}\label{mx2}
  F^{0 i}{}_{,i}=F^{i 0}{}_{,i}=0~,
\end{equation}

\begin{equation}\label{mx22}
  F^{j 0}{}_{,0}+F^{j i}{}_{,i}=0~,
\end{equation}
or
$$(h_+-1)E_{1,1}- (1+h_+)E_{2,2}-E_{3,3}=0~,   $$
$$ \left((1-h_+)E_1\right)_{,0}+B_{3,2}-B_{2,3}=0~,   $$
$$ \left((1+h_+)E_2\right)_{,0}-B_{3,1}+B_{1,3}=0~,   $$
$$  E_{3,0}+B_{2,1}-B_{1,2}=0~.$$
The Maxwell equations (\ref{maxequation2})
$$F_{0i,j}+F_{j0,i}+F_{ij,0}~=0~,$$
$$F_{ij,k}+F_{ki,j}+F_{jk,i}~=0~,$$
reduce to
$$ E_{1,2}-E_{2,1}+B_{3,0}=0,   $$
$$ E_{1,3}-E_{3,1}-\left(\left(1+h_+\right)B_2\right)_{,0}=0~,   $$
$$ E_{2,3}-E_{3,2}+\left(\left(1-h_+\right)B_1\right)_{,0}=0~,   $$
$$ B_{3,3}+\left(1+h_+\right)B_{2,2}+\left(1-h_+\right)B_{1,1}=0~.$$
The above equations can be rewritten into a compact form
\begin{equation}\label{mx3}
  \left\{
     \begin{array}{ll}
       \nabla \cdot \emph{\textbf{E}}=h_+\left(E_{1,1}-E_{2,2}\right)~,  \smallskip \\
       \dfrac{\partial{\emph{\textbf{E}}}}{\partial t}-\dfrac{\partial{\emph{\textbf{h}}}}{\partial t}\cdot
       \emph{\textbf{E}}+\nabla \times \emph{\textbf{B}}=0~,  \smallskip \\
       -\nabla \times \emph{\textbf{E}}+\dfrac{\partial{\emph{\textbf{B}}}}{\partial t}=
       \dfrac{\partial{\emph{\textbf{h}}}}{\partial t}\cdot \emph{\textbf{B}}~,  \smallskip \\
       \nabla\cdot \emph{\textbf{B}}=h_+\left(B_{1,1}-B_{2,2}\right)~,
     \end{array}
   \right.
\end{equation}
where $\emph{\textbf{E}}$ and $\emph{\textbf{B}}$ are  electric
field  and magnetic field vector in column matrix form and
$\emph{\textbf{h}}$ is a square matrix:
\begin{equation}
\emph{\textbf{h}}=
  \left(
  \begin{array}{ccc}
    h_+ & 0 & 0 \\
    0 & -h_+ & 0 \\
    0 & 0 & 0 \\
  \end{array}
\right)~,
\end{equation}
$\nabla$ is the gradient operator in 3-dimensional flat space.
These are the deformed Maxwell equations in the background of
GW in a flat space-time approximation.

We try to setup the electromagnetic wave
equations from the deformed Maxwell equations. First, making a time
derivative at both sides of the second equation \eqref{mx3} and using
the third equation, we get
\begin{align}
 \frac{\partial^2{\emph{\textbf{E}}}}{\partial t^2}-\nabla^2\emph{\textbf{E}} & =
 \frac{\partial^2{\left(\emph{\textbf{h}}\cdot \emph{\textbf{E}}\right)}}{\partial t^2}-\nabla\left(\nabla \cdot
 \emph{\textbf{E}}\right)-\nabla \times \frac{\partial{\left(\emph{\textbf{h}} \cdot \emph{\textbf{B}}\right)}}{\partial t}~.
\end{align}
The ratio of the frequency of GW and that of photons
is very small and can be used as a perturbation parameter.  Then,
\begin{align}
 &\frac{\partial^2{\emph{\textbf{E}}}}{\partial t^2}-\nabla^2\emph{\textbf{E}}= \emph{\textbf{h}}\cdot\frac{\partial^2{\emph{\textbf{E}}}}{\partial t^2}-h_+\nabla\left(E_{1,1}-E_{2,2}\right)-
\nabla\times\left(\emph{\textbf{h}}\cdot \left(\nabla \times
\emph{\textbf{E}}\right)\right)~.
\end{align}
It can be rewritten as
\begin{align}
  \left(1-\emph{\textbf{h}}\right)\cdot&\frac{\partial^2{\emph{\textbf{E}}}}{\partial t^2}-\nabla^2\emph{\textbf{E}} = -h_+\nabla\left(E_{1,1}-E_{2,2}\right)-
\nabla\times\left(\emph{\textbf{h}}\cdot \left(\nabla \times
\emph{\textbf{E}}\right)\right).
\end{align}

In components, the deformed electromagnetic wave equations are of the
form
\begin{equation}\label{compform}
  \left\{
     \begin{array}{ll}
       \left(1-h_+\right)\dfrac{\partial^2{E_1}}{\partial t^2}-\nabla^2E_1=-h_+\left(E_{1,11}-E_{2,21}\right)
       -h_+\left(E_{1,33}-E_{3,13}\right),   \smallskip \\
       \left(1+h_+\right)\dfrac{\partial^2{E_2}}{\partial t^2}-\nabla^2E_2 =-h_+\left(E_{1,12}-E_{2,22}\right)
       -h_+\left(E_{3,23}-E_{2,33}\right),  \smallskip  \\
       \dfrac{\partial^2{E_3}}{\partial t^2}-\nabla^2E_3=
h_+\left(E_{3,22}-E_{3,11}\right) ~.
     \end{array}
  \right .
\end{equation}
The similar wave equations for $\emph{\textbf{B}}$ can be obtained in the same way.

When the electromagnetic wave propagates along $x$ axis, $E_1=0$.
Then, the first equation of \eqref{compform} is the identity $0=0$ in the leading order.  The latter two equations
read
\begin{equation}\label{compform2}
    \left\{
     \begin{array}{ll}
       \left(1+h_+\right)\dfrac{\partial^2{E_2}}{\partial t^2}-\nabla^2E_2=h_+E_{2,22}-h_+\left(E_{3,23}-E_{2,33}\right),  \smallskip \\
       \dfrac{\partial^2{E_3}}{\partial t^2}-\nabla^2E_3=h_+\left(E_{3,22}-E_{3,11}\right) ~.
     \end{array}
  \right.
\end{equation}
Since the right hand side of the first equation of (\ref{compform2}) is
\begin{align}
 & h_+E_{2,22}+h_+[(E_{1,1}+E_{2,2}+h_+E_{2,2}-h_+E_{1,1})_{,2}+E_{2,33}]
   \notag \\
   & = h_+E_{2,22}+h_+\left(E_{2,22}+E_{2,33}\right) \notag \\
   & = h_+E_{2,22}+h_+\left(\nabla^2E_2-E_{2,11}\right)~,
\end{align}
the first equation of (\ref{compform2}) reduces to
\begin{equation}
\left(1+h_+\right)\frac{\partial^2{E_2}}{\partial t^2}-\left(1+h_+\right)\nabla^2E_2  =h_+\left(E_{2,22}-E_{2,11}\right)~.
\end{equation}
The leading order of the perturbation of the electromagnetic wave's equation is
\begin{equation}
\frac{\partial^2{E_2}}{\partial t^2}-\nabla^2E_2  =h_+\left(E_{2,22}-E_{2,11}
\right)~.
\end{equation}
For the electromagnetic wave propagating along the $x$ axis, $E_{2,22}=0$ , $E_{2,33}=0$, $E_{3,22}=0$ , $E_{3,33}=0$.  Hence,
\begin{align}\label{speed1a}
  \frac{\partial^2{E_2}}{\partial t^2}-\left(1-h_+\right)E_{2,11}&=0~.
\end{align}
The similar deduction gives rise to
\begin{align}\label{speed1b}
   \frac{\partial^2{E_3}}{\partial t^2}-\left(1-h_+\right)E_{3,11}&=0~.
\end{align}
\eqref{speed1a} and \eqref{speed1b} are just the deformed form of the wave
equation for electric field.  The equations \eqref{speed1a} and \eqref{speed1b}
gives the coordinate speed of the electromagnetic wave in the background of
GW,
\begin{equation}
    v_x=\sqrt{1-h_+\left(t\right)}\approx 1-\frac{h_+}{2}~.
\end{equation}
Similarly, one can obtain that
\begin{equation}
 v_y=\sqrt{1+h_+\left(t\right)}\approx 1+\frac{h_+}{2}.
\end{equation}
These mean that the photon's coordinate speeds vary with the
perturbation of space-time, which is caused by GW.
The similar result has been obtained in other ways (say, in \cite{MRSWD}).

Suppose that the electromagnetic wave spend $T'$ and $T$ to propagate from $A$ to $B$ and then
back to $A$ when there is and there is no GW, respectively.
The time difference $\Delta T'$ due to GW satisfies
\begin{eqnarray}\label{DeltaTUsingMax}
  2L &=& \int_{t_0}^{t_0+2T+\Delta T'}v\od t = \int_{t_0}^{t_0+2T+\Delta T'}\left(1-\frac{1}{2}\Re[e_{11}e^{-i (2\pi f
  t+\phi)}]\right)\od t~.
\end{eqnarray}
Again, set $f=\frac{a}{2L}$ and $t_0=0$.  Eq.(\ref{DeltaTUsingMax}) results in
\begin{equation}
  \Delta T'=\frac{e_{11}L}{2a\pi}\left(\sin\left(\frac{a\pi(\Delta T'+2L)}{L}
  +\phi\right)-\sin\phi\right)~.
\end{equation}
Its leading terms are again given by Eq. \eqref{3.4}
because $\Delta T'$ is much smaller than $L$.  Obviously,
when the dimensionless frequency $a = 1,2,3...~$, $\Delta T' = 0$, which means that the detector has zero response.

\section{Gravitational wave vs swing of end mirror}

In the above, only the ideal cases are discussed.  Practically, an interferometer
will undergo various vibrations.  Whether it is possible to distinguish the GW
from the swing of the mirrors is an important problem in the detection of GW.
In the following, we shall show the difference between the effect of GW
and the swing of a mirror.

For simplicity, we consider the case that the mirror $A$ is fixed and the mirror $B$
undergoes a harmonic oscillation $\Re[Ae^{-i(2\pi ft +\phi)}]$, where $f$ is now the
frequency of the vibration and $\phi$ is the initial phase of the $B$'s oscillation
when the light leaves the mirror $A$.
Let $\tau$ be the traveling or propagating time of light for one-way trip when the mirror $B$ undergoes an oscillation.  Then,
\begin{equation}
  2\tau-2L=\Re\left[2\left(Ae^{-i(2\pi f t +\phi)}\right)\right]~.
\end{equation}
Denote the difference of the traveling or propagating time for the light's one-way trip with and without
end mirror's swing by $\Delta T$. We let a photon leaves mirror $A$ at $t=0$ and consider the photon with  initial phase  $\phi$. Then, in the leading approximation,
\begin{equation}\label{vibrating}
  \Delta T=\Re[Ae^{-i(2\pi fL  +\phi)}]~.
\end{equation}

$\Delta T =0$ requires that
\begin{gather}
\Re[e^{-i(2\pi f L +\phi)}]=0~,\\
2\pi fL+\phi=\left(\frac{1}{2}+n\right)\pi~,\\
 f=\frac{n}{2L}+\frac{1}{2L}\left(\frac{1}{2}-\frac\phi{\pi}\right)~.
\end{gather}
Obviously, the initial phase of the mirror's swing
will influence the difference of the traveling time.  Since a continuous laser
is used, the initial phases are uniformly distributed, equivalently.
In other words, no matter what frequency the mirror $B$ swings at, there
are always some photons who can `feel' the swing.
In contrast, the photons cannot `feel' the GW with frequency
$f=\frac n{2L}$.

\begin{figure}
  \centering
  \includegraphics[width=12cm]{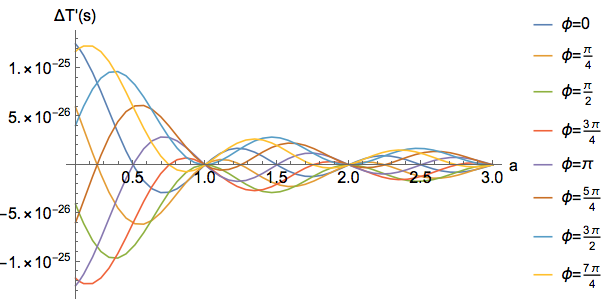}\\
  \caption{Time difference of photon for a round trip from $A$ to $B$ and back to $A$
  as a function of frequency of GW. Parameters are chosen as $t_0=0$,
  $c=3\times 10^8$m/s,
  $e_{11}=10^{-20}$ and $L=4000$m. The time difference also depends on the initial phase of
  GW.  The different color curves represent the different initial phases.
  }\label{figure1}
\end{figure}
Figure. 1 presents the plot of $\Delta T'$ due to a GW, given by Eq.
\eqref{3.4}.  (The parameters are chosen as 
$e_{11}=10^{-20}$ and $L=4000$m.)  The different color denotes the different initial
phase $\phi$ of GW.  In the figure, the 8 curves correspond to
$\phi=0,\frac{\pi}{4},\frac{\pi}{2},\frac{3\pi}{4},\pi,\frac{5\pi}{4},\frac{3\pi}{2}$,
and $\frac{7\pi}{4}$, respectively.
From the figure, we can see that the time difference of photon motions always vanishes when
the frequency of a GW is $\frac{n}{2L}$.
It means that a photon spends the same time to travel for a round trip in the background of specific GW with frequency $\frac{n}{2L}$ as in a flat space-time.
From Figure.\ref{figure1}, we can also see that the initial phase of
GW affects the readout in the measurement of GW.  The envelop has the form of sinc\,$x=\frac{\sin x}x$.

\begin{figure}
  \centering
  \includegraphics[width=12cm]{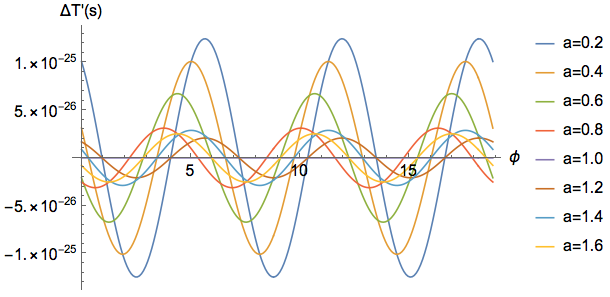}\\
  \caption{The initial phase dependence of the measurement for GW.
  The frequency of GW takes as $f=\frac{a}{2L}$, and $a=0.4,0.6,0.8,1.0,1.2,1.4,1.6$.}\label{figure2}
\end{figure}

Figure.\ref{figure2} shows the dependence of time difference of photon motions on
the initial phase of GW for specific frequencies. It should be noticed
that the curve corresponding with $f=\frac{1}{2L}$ (i.e. $a=1$) coincides
with the abscissa axis exactly.  It means that the photon cannot `feel' the GW with $f=\frac{1}{2L}$ no matter what the initial phase is.

\begin{figure}
  \centering
  \includegraphics[width=12cm]{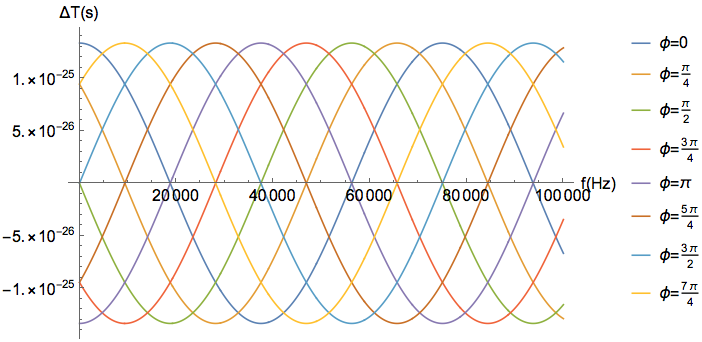}\\
  \caption{Frequency dependence of the time difference of photon
motions caused by the mirror's vibration.  Parameters are taken as $L=4000$m,
$c=3\times10^8$m/s, $A=4\times10^{-17}$m, again.  For specific values of the initial phase,
  $\phi=0,\frac{\pi}{4},\frac{\pi}{2},\frac{3\pi}{4},\pi,\frac{5\pi}{4},
  \frac{3\pi}{2},\frac{7\pi}{4}$, are depicted in different colors. }\label{figure3}
\end{figure}
\begin{figure}
  \centering
  \includegraphics[width=12cm]{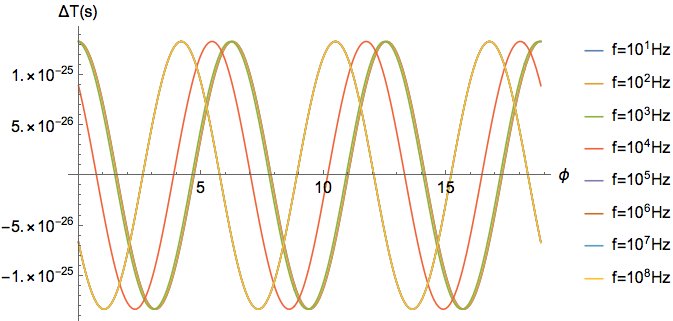}\\
  \caption{Initial phase dependence of the detector's response. Parameters are taken as $L=4000$m,
$c=3\times10^8$m/s, $A=4\times10^{-17}m$.  The different vibration frequencies are depicted in
different colors.}\label{figure4}
\end{figure}

In the case of the mirror's swing, the time difference of photon motions
is
\begin{equation}\label{vi}
    \Delta T=A \cos\left(2\pi fL+\phi\right)~.
\end{equation}
We plot the frequency dependence of the time difference of photon
motions caused by the mirror's swing in Figure.\ref{figure3}.
Form the picture, we can see that the responses of the interferometer to
different frequencies are almost the same.  In this case, the time difference of
photon motions also depends on the initial phase when this
photon leaves the mirror $A$.  The initial phase dependence is shown in Figure.\ref{figure4}.

\section{\label{sec:level1}Conclusions and remarks}

In this paper, the photon motions in the background of GW are re-analyzed in both
geometrical and wave optics. In geometrical optics, the path of photons in 3-dimensional
space is obtained by using the Fermat principle.  The geodesic equations are
re-produced and the photon paths are just the null geodesics in general relativity.
The proper distance between the two mirrors and
thus the traveling time that photons move in-between change with GW.
In wave optics, the deformed Maxwell equations and thus the wave equations for electric
fields are presented.  The wave equations show that the coordinate speed $v(t)$ of an
electromagnetic wave moving between the two mirrors is modified by GW.  The two methods
give the exactly same equation \eqref{3.4} for the time difference influenced by GW.

Both GW and the mirror swings result in the variations of the proper distance between
the two mirrors.  However, the characteristics of the variations are different.  For a gravitational wave, the
amplitude of $\Delta T'$ varies with the frequency of the GW, which is given by the
envelop in Figure. 1.  The property of the envelope has been reported in literature (see, for example, \cite{SC}, \cite{CPC}, \cite{MRSWD} etc.).  For the mirror swing, however, the
amplitude of $\Delta T$ is independent of the frequency of swings.
The difference can be easily seen from
the figures.  The reason of the difference is as follows.  A swinging mirror only affects
the reflecting point of light, but does not affect the propagation of light in the arm.
On the contrary, a GW affects the propagation of light in the arm everywhere.
Therefore, the swinging mirror cannot mimic a gravitational
wave very well, which is used in the previous studies on the instruments and
calibration \cite{MRSWD}, \cite{DynRes}, \cite{Readout}, \cite{Character}, \cite{TestMass}, \cite{cqg}, \cite{calibration}.

The above discussions are just for the motion of photons in a single arm.  All the discussions
are also valid for the motion of photons in the other arm.  The only difference is that the phase variation in the two arms have different polarity due to the quadruple nature of the GW.
The total effect should be the sum for two arms. If the time differences in both
arms vanish for some specific frequency of GWs, then the detector has no response
to these GWs.  For the swinging mirrors, the detector may provide zero-response
only when the two arms vibrates in synchronization.

Finally, the time difference in a Michelson interferometer is discussed in the present paper.
The detectors, like LIGO, Virgo, are all Michelson-Pabry-Perot interferometers.
The analysis for a Michelson-Pabry-P\'erot interferometer will be discussed elsewhere.

\section*{Acknowledgments}
The project is supported by National Natural Science Foundation of China under the grants 11275207, 11375203, 11690022 and 11675182 and by the Strategic Priority Research Program
of the Chinese Academy of Sciences ¡°Multi-waveband Gravitational Wave Universe¡± (Grant
No. XDB23040000).


\bigskip

\section*{References}
\begin{thebibliography}{10}
\bibitem{LVC00} B. P. Abbott, et al, 2016 Observation of gravitational waves from a binary black hole merger, \textit{Phys. Rev. Lett.} {\bf 116} 061102
\bibitem{LVC10}B. P. Abbott, et al, 2016
GW151226: Observation of Gravitational Waves from a 22-Solar-Mass Binary Black Hole Coalescence
\textit{Phys. Rev. Lett.} {\bf 116} 241103

\bibitem{Black} E. D. Black and R. N. Gutenkunst, 2003
An Introduction to Signal Extration in Interferometric Gravitational Wave Detectors,
\textit{Am. J. Phys.} {\bf 71} 365

\bibitem{MatchFilter} B. Allen, W. G. Anderson, P. R. Brady, D. A. Brown, and J. D. E. Creighton, 2012
FINDCHIRP: An algorithm for detection of gravitational waves from inspiraling compact binaries
\textit{Phys. Rev. D} {\bf 85} 122006.

\bibitem{aLIGO} J. Aasi et al., 2015
Advanced LIGO,
\textit{Class. Quant. Grav.} {\bf32} 074001

\bibitem{SC} Z. Chang, C.-G. Huang, Z.-C. Zhao, 2016
Propagation effect of gravitational wave on detector response.
\textit{Sci. China--Phys. Mech. Astro.} {\bf 59} 100421

\bibitem{CPC} Z. Chang, C.-G. Huang, Z.-C. Zhao, 2017
Is GW151226 a really signal of gravitational wave?
\textit{Chin. Phys. C} {\bf 41} DOI:10.1088/1674-1137/41/2/025001.

\bibitem{MRSWD}J. Mizuno, A. R\"udiger, R. Schilling, W. Winkler, K. Danzmann, 1997
Frequency response of Michelson- and Sagnac-based interferometers
\textit{Optics Commun.} {\bf 138} 383-393.

\omits{\bibitem{Black} E. D. Black and R. N. Gutenkunst, 2001
Am. J. Phys. 69: 79-87 }

\bibitem{DynRes} M. Rakhmanov, R. L. Savage, D. H. Reitze, et al. 2002
Dynamic resonance of light in Fabry¨CPerot cavities[J].
\textit{Phys. Lett. A} {\bf 305} 239-244

\bibitem{Readout}J. Markowitz, R. Savage, and P. Schwinberg. 2003
Development of a Readout Scheme for High Frequency Gravitational Waves.
\textit{LIGO Technical Document} T030186-00-W

\bibitem{Character} M. Rakhmanov, F. Bondu, O. Debieu, et al. 2004
Characterization of the LIGO 4 km Fabry¨CPerot cavities via their high-frequency dynamic responses to length and laser frequency variations[J].
\textit{Class. Quantum Grav.} {\bf 21} S487

\bibitem{TestMass}M. Rakhmanov, 2005
Response of test masses to gravitational waves in the local Lorentz gauge
\textit{Phys. Rev. D} {\bf 71} 084003

\bibitem{cqg} M. Rakhmanov, J. D. Romano, and J. T. Whelan, 2008
High-frequency corrections to the detector response and their effect on searches for gravitational waves.
\textit{Class. Quantum Grav.} {\bf 25} 184017

\bibitem{calibration}B. P. Abbott, LIGO Scientific Collaboration, 2016
Calibration of the Advanced LIGO detectors for the discovery of the binary black-hole merger GW150914[J]
\textit{arXiv preprint} (arXiv:1602.03845)

\bibitem{Mizuno}J. Mizuno, 1995
Comparison of optical configurations for laser-interferomettric gravitational-wave detectors
\textit{Doctoral dissertation, Hannover University.}

\bibitem{Landau} L. D. Landau and E. M. Lifshitz, The classical theory of fields, 1980
4 edition, \textit{Butterworth-Heinemann, Amsterdam.}

\bibitem{Thorne}K. S. Thorne, 1987
Gravitational radiation in Three hundred years of gravitation, ed by S. W. Hawking and W. Israel,
\textit{Cambridge University Press, Cambridge.}


\end{thebibliography}
\end{document}